\pdfoutput=1

\documentclass{article}
\usepackage[utf8]{inputenc}

\usepackage{tikz}
\usetikzlibrary{backgrounds,shapes,arrows,positioning,calc,decorations,fit}
\usetikzlibrary{arrows.meta,decorations.markings}

\usepgflibrary{decorations.markings}

\usepackage{pgfplots}
\pgfplotsset{compat=1.12}

\usepackage{graphicx}

\usepackage{etoolbox}
\apptocmd{\sloppy}{\hbadness 10000\relax}{}{}

\usepackage{slashbox}

\newcommand{\etal}{\textit{et al}.}
\newcommand{\ie}{\textit{i}.\textit{e}.,}
\newcommand{\eg}{\textit{e}.\textit{g}.,}

\usepackage{enumitem}

\usepackage{booktabs}

\usepackage{comment}

\excludecomment{thesisNavigation}

\newcommand{\chapterPaper}[2]{%
  \ifnum\pdfstrcmp{\cond}{chapter}=0
    \ifnum\pdfstrcmp{}{#1}=0\unskip\else#1\fi%
  \else
    \ifnum\pdfstrcmp{\cond}{paper}=0
      \ifnum\pdfstrcmp{}{#2}=0\unskip\else#2\fi \fi %
  \fi\ignorespaces}

\def\cond{paper}

\usepackage{hyperref}

\providecommand{\keywords}[1]
{
  \small	
  \textbf{\textit{Keywords---}} #1
}

\title{CRC: FULLY GENERAL MODEL OF \\ \textbf{C}ONFIDENTIAL \textbf{R}EMOTE \textbf{C}OMPUTING
\thanks{This document is the sixth chapter of DPhil (PhD) thesis of the first author.}}

\author{Kubilay Ahmet K\"{u}\c{c}\"{u}k\thanks{Corresponding author. Reachable at kucuk@acm.org after Oxford affiliation.}, Andrew Martin \\
Department of Computer Science \\
\textit{University of Oxford}\\
kucuk,andrew.martin@cs.ox.ac.uk
}
\date{Ver. April 2023 Release\thanks{Based on research since September 2015. The main content is written in 2017-2019. \\An earlier version is released by Dec 2020.}}

\usepackage{imakeidx}

\usepackage[automake,acronym,toc]{glossaries}

\newacronym{crc}{CRC\index{Confidential Remote Computing}}{Confidential Remote Computing\index{Confidential Remote Computing}}
\newacronym{drtm}{DRTM\index{Dynamic Root of Trust for Measurement}}{Dynamic Root of Trust for Measurement\index{Dynamic Root of Trust for Measurement}}
\newacronym{pal}{PAL\index{Piece of Application Logic}}{Piece of Application Logic\index{Piece of Application Logic}}
\newacronym{pcr}{PCR\index{Platform Configuration Registers}}{Platform Configuration Registers\index{Platform Configuration Registers}}
\newacronym{pts}{PTS}{Platform Trust Services\index{Platform Trust Services}}
\newacronym{rtm}{RTM\index{Root of Trust for Measurement}}{Root of Trust for Measurement\index{Root of Trust for Measurement}}
\newacronym{rtr}{RTR\index{Root of Trust for Reporting}}{Root of Trust for Reporting\index{Root of Trust for Reporting}}
\newacronym{rts}{RTS\index{Root of Trust for Storage}}{Root of Trust for Storage\index{Root of Trust for Storage}}
\newacronym{sgx}{SGX\index{Software Guard eXtensions}}{Software Guard eXtensions\index{Software Guard eXtensions}}
\newacronym{srtm}{SRTM\index{Static Root of Trust for Measurement}}{Static Root of Trust for Measurement\index{Static Root of Trust for Measurement}}
\newacronym{tacb}{TaCB\index{Trustable Computing Base}}{Trustable Computing Base\index{Trustable Computing Base}}
\newacronym{tcb}{TCB\index{Trusted Computing Base}}{Trusted Computing Base\index{Trusted Computing Base}}
\newacronym{tee}{TEE\index{Trusted Execution Environment}}{Trusted Execution Environment\index{Trusted Execution Environment}}
\newacronym{tpm}{TPM\index{Trusted Platform Module}}{Trusted Platform Module\index{Trusted Platform Module}}
\newacronym{txt}{TXT\index{Trusted eXecution Technology}}{Trusted eXecution Technology\index{Trusted eXecution Technology}}

\newacronym{edl}{EDL\index{Enclave Description Language}}{Enclave Description Language\index{Enclave Description Language}}

\newacronym{tcg}{TCG\index{Trusted Computing Group}}{Trusted Computing Group\index{Trusted Computing Group}}

\newacronym{qe}{QE\index{Quoting Enclave}}{Quoting Enclave\index{Quoting Enclave}}

\newacronym{ho}{HO\index{Hardware Owner}}{Hardware Owner\index{Hardware Owner}}

\newacronym{ao}{AO\index{Algorithm Owner}}{Algorithm Owner\index{Algorithm Owner}}

\newacronym{do}{DO\index{Data Owner}}{Data Owner\index{Data Owner}}

\newacronym{ief}{IEF\index{Internal Enclave Functions}}{Internal Enclave Functions\index{Internal Enclave Functions}}

\newacronym{epm}{EPM\index{Early Private Mode}}{Early Private Mode\index{Early Private Mode}}

\newacronym{pief}{PIEF\index{Public Internal Enclave Functions}}{Public Internal Enclave Functions\index{Public Internal Enclave Functions}}

\newacronym{sief}{SIEF\index{Secret Internal Enclave Functions}}{Secret Internal Enclave Functions\index{Secret Internal Enclave Functions}}

\newacronym{ssief}{SSIEF\index{Serialised Secret Internal Enclave Functions}}{Serialised Secret Internal Enclave Functions\index{Serialised Secret Internal Enclave Functions}}

\newacronym{pcl}{PC Loader\index{Protected-Code Loader}}{Protected-Code Loader\index{Protected-Code Loader}}

\makeglossaries

\begin{document}

\maketitle

\begin{abstract}
Digital services have been offered through remote systems for decades. 
The questions of how these systems can be built in a trustworthy manner and how their security properties can be understood are given fresh impetus by recent hardware developments, allowing a fuller, more general, exploration of the possibilities than has previously been seen in the literature. 
Drawing on and consolidating the disparate strains of research, technologies and methods employed throughout the adaptation of confidential computing, we present a novel, dedicated \acrlong{crc} model. 
\acrshort{crc} proposes a compact solution for next-generation applications to be built on strong hardware-based security primitives, control of secure software products' trusted computing base, and a way to make correct use of proofs and evidence reports generated by the attestation mechanisms. 
The \acrshort{crc} model illustrates the trade-offs between decentralisation, task size and \emph{transparency overhead}. 
We conclude the chapter with six lessons learned from our approach, and suggest two future research directions explained.

\end{abstract}


\clearpage

\tableofcontents

\clearpage
    
\printglossary[type=\acronymtype,title={List of Acronyms}, toctitle=List of Acronyms]


\clearpage




\section{What is Confidential Remote Computing?}

Computer programs contain computations. 
Simply put, the function $f(x)=xa+b$, a computation, or a NAND gate, consists of two elements, the data (input) and the algorithm (operation).
These computations can require confidentiality for their assets.
The basis of any confidentiality is one party's desire for protection from another's infringement on their assets.
Although a computation can run locally, it turns into a remote computation between these two parties involved.
This is why \emph{Confidential Computing\footnote{\url{https://confidentialcomputing.io/}}} cannot be conceived without the key aspect of remoteness.

The problem we identify and response in \emph{\acrlong{crc}} has been around for decades. 
In fact, we shall discuss in Section~\ref{sec:history} that it has been studied throughout the history of inter-connected computations.
We shall highlight the development of old and new aspects of \emph{\acrlong{crc}}.
Over time, new actors introduced new approaches, evolving requirements and corresponding security solutions.
Thus, the multiplicity of actors in a computation poses a challenge for any emerging control decentralisation.
The main issue is decentralisation of the computation, our solution approaches to this problem through its participants.

In this \chapterPaper{chapter}{paper}, we attempt to consolidate the disparate research approaches into one \emph{Confidential Remote Computing} model, from past to future.
We begin by explaining the participating entities in chronological order, divided into five subsequent stages.
The model then is structured in three main sections.
First, we begin with the role of hardware technologies in \emph{Confidential Remote Computing}.
Then we introduce our model, demonstrating the orthogonal research structured by each entity.
Lastly, we demonstrate the key trade-offs providing an overview of the deployed systems.

\subsection{The Problem Statement}\label{sec:CRCproblemDef}

The problem of \acrshort{crc} can be expressed in various means. 
For simplification and inter-disciplinary understanding, we explain the problem by using a classic example with digitalised companies, \eg~take two insurance companies.
Both of these companies promise to perform advertised computational tasks remotely with your private information and promise you to delete all the information stored.
One of them has certified infrastructure, isolated and confidential execution, and minimised software stack.
The other one is infected by advanced malware, it may perform some hidden operations (\eg~selling your data off in the background).
Can you tell the difference between the two from your home computer?
What proof could there be that you are communicating with the right insurance company?
Similarly, consider a bank, a block-chain infrastructure, your government, and many other mission-critical online services you are using.
How do you know that the behaviour of the remote system is, as promised, trustworthy, to whatever end you wish to employ it?

The service quality of a business depends on the transparency it offers.
Being able to offer true trustworthy service is not only an added value to products, it is seen as a basic requirement by users.
Gaining trust may come through formal validation and verification methods (\ie~checking if a piece of code matches its specifications) in relatively small systems.
In contrast, our focus in this \chapterPaper{chapter}{paper}~shall be placed on the \emph{Confidential Computing} paradigm, in order to help us understand how the root of trust in hardware can be translated to humans through software components.
For example, users can make sure that an election was counted correctly, cast-as-intended (\eg~towards universal verifiability). 
Alternatively, a user can reverse-monitor\footnote{Reverse-monitor; the action of observing the observer entity. \ie~if monitoring takes place in one direction from a company towards their users, these users watch the company's actions in reverse direction.} an online advertisement company to verify it does not invade her privacy.

\subsection{The Ambiguity of Trust in Naming Convention}\label{sec:ambiguityofTrust}

\begin{table}[hpt]
    \centering
        \begin{tabular}{l|l}
        \hline
        Trustworthy Systems    & Used for security notions by verification \\ \hline
        Trustworthy Computing  & Used for security notions by measurement  \\ \hline
        Trusted Computing  &  Older term from DoD's Orange Book  \\ \hline
        Trustworthy Remote Computing & Entities with TPM/Enclave computing \\ \hline
        Confidential Computing   &  Used by industry, Microsoft in Azure \\ \hline
        \acrlong{crc}  & Term we model in the rest of the \chapterPaper{chapter}{paper}\\ \hline
        Secure Remote Computing & Term for the ideal case  \\ \hline
        \end{tabular}
        \caption{Overview of the terms with closely related meanings in the field.}\label{tab:namingConvention}
\end{table}

Trustworthiness implies integrity guarantees among other security notions.
Integrity can be observed through measurement or with the means of verification.
Through software verification, other guarantees specified such as confidentiality can be obtained. 
Despite the presence of other security guarantees through verification, however, integrity is the key notion to trustworthiness.
In table~\ref{tab:namingConvention}, we summarised these concepts.
Although confidentiality properties can be derived by measurement or verification, the sub-notions such as privacy, data-secrecy, algorithm-secrecy, require additional attention based on threat modelling.
Physical access, \eg~through side-channel leakages, in a system can cause confidentiality issues.
For example, software verification can guarantee that a piece of information is kept confidential throughout the life-cycle of the execution, avoiding software bugs leaking the information.
In this \chapterPaper{chapter}{paper}, we focus on \emph{\acrlong{crc}}, a concept which requires both strong integrity (due to remote execution) and strong confidentiality guarantees.
The older term \emph{trusted} was used in the Department of Defence's orange book~\cite{latham1985department}, however, the current understanding replaces it with \emph{trustworthy} where the users can evaluate the pieces of evidence present and make a trust decision.
There are instances of the term \emph{trustworthy systems}\footnote{Trustworthy Systems used by research groups. \url{https://ts.data61.csiro.au}} used to imply \emph{trust by verification}.
\emph{Trustworthy computing} is often used to describe the field of trusted computing, where the pieces of evidence are generated with/from hardware \emph{root of trust} primitives.
There are secure systems built~\cite{Kucuk2016} with hardware technologies (such as \acrshort{tee}-based or \acrshort{tpm}-based, etc.), where these systems address the security problems of \emph{trustworthy remote computing}.
The industry consortium of confidential computing also uses a similar naming convention, however, it implies a product name/group in cloud computing.
Some researchers, \eg~Johnson~\etal~\cite{sgxMutliSocket} and Russinovich~\etal~\cite{10.1145/3454122.3456125}, emphasise the cloud use of confidential computing, rebranding it as \emph{confidential cloud computing}.
Finally, the ideal case of \emph{secure remote computing~\cite{costan2016intel}} is an unsolved problem.
While there may be small-scale proof-of-concept studies, \emph{secure remote computing} is difficult to perform as a many-party, high-scale, generic computation.

\subsection{Related Work}

Confidential Computing Consortium's July 2020 white paper~\cite{cccWhitePaper2020} shows the benefits of hardware assistance in computations, \eg~the encryption-in-use for data.
It includes a good comparison of the cryptographic methods (\ie~from homomorphic encryption standardisation) and the trusted hardware. 
For example, in homomorphic encryption schemes, publicly known code is mutually agreed on by the participants (data providers), therefore, they do not provide code secrecy to hide the private algorithms.
The paper shows that trusted execution environments (\acrshort{tee}s) can provide attestability, code confidentiality, and better programmability on top of the existing guarantees of homomorphic encryption.
The code confidentiality with \acrshort{tee}s is described as \emph{requiring further work}, though until now, secret-code execution in \acrshort{tee}s has been studied in recent years~\cite{kuccuk2019managing, silva2017dynsgx, bauman2018sgxelide}, and a \acrfull{pcl} is also an integrated feature in the Intel \acrshort{sgx}'s SDK.
Another short introduction~\cite{rashid2020rise} to confidential computing states how widely the hardware-assistance is adopted by major industry vendors.
An abandoned patent~\cite{frank2015system} and a valid patent~\cite{naganuma2016confidential} state similar concepts.
However, Naganuma's patent~\cite{naganuma2016confidential} proposes confidential computing through homomorphic encryption, with no use of hardware-assisted execution technologies.
To the best of our knowledge, our work is the first attempt to model \emph{\acrlong{crc}} in academia.

\subsection{Contributions}

This \chapterPaper{chapter}{paper}~expresses the most comprehensive model of \emph{\acrlong{crc}}. 
The \emph{\acrlong{crc}} paradigm brings new opportunities to the computing world.
Our model aims to provide a structured view of the technologies and methods behind the digitalisation trend of the concept of trust.
Throughout the \chapterPaper{chapter}{paper}, we make the following contributions with the novel \emph{\acrlong{crc}} model:

\begin{enumerate}[noitemsep]
    \item \acrshort{crc} resolves the ambiguity of \emph{trust} in relevant domains, and we connect the former understanding of \emph{trust} in computations (local and certificate-based) to future computing models (with separated three entities; cloud providers, data providers and algorithm providers).
    \item We provide an extensive analysis on how \emph{digital trust} can be derived from alternative methods such as micro-kernels, software-based attestation, and verification technologies, followed by the use of the hardware enclaves in the grid, edge and fog computing use cases.
    \item We consolidate the \acrshort{crc} model in the X-chart in Figure~\ref{figure:domains_X_Chart} in three domains; hardware, attestation and development, presented in Section~\ref{sec:xchart}.
    Each of these domains is surveyed and systematised with applied techniques and methods in the literature.  
    \item We present the trade-offs in the \acrshort{crc} model, between the larger task size, more decentralisation cost and more transparency overhead.
\end{enumerate}

We conclude the \chapterPaper{chapter}{paper}~with a unified architecture utilising the analysed technologies, and present the lessons learned alongside the future research directions.

\begin{thesisNavigation}

\subsection{Position in the Thesis}
This \chapterPaper{chapter}{paper}~provides the most generic form of the insights and the systems developed in Part II (with Public Code and Private Data) and Part III (with Private Code and Private Data).
The importance of the systematisation presented is that it provides a novel model for the evolution of trust in the history of \emph{Confidential Remote Computing}.
It shows how the perception of \emph{trust} is adopted through the co-evolution of the requirements and solutions brought by new stakeholders of the computations.
The final abstract model of the unified architecture is presented in the paper as a basis for future research as suggested in Part V of the thesis.

\end{thesisNavigation}

\subsection{Position of \emph{\acrlong{crc}} in Five Stages of Computing History}
\label{sec:history}
Similar challenges to ones found in \emph{\acrlong{crc}} have been around in computing history.
Trust and confidentiality challenges have been around for decades, never fully solved, and will likely continue to evolve with new advances in computing by new participants.
In this brief presentation of \emph{\acrlong{crc}}, we connect the past and future of remote computation.
There is a progress in these stages on how the computations were carried.
This list is not to say that every single computer followed them, \textbf{we justify this list with the advancement of the technologies.}
Once a technology became available, \eg~PKI, we move into the next stage.
For example, with the introduction of hardware-assisted memory encryption, attestation and isolation features, we move into the fifth stage of computations where algorithms can be hidden while offering services in commodity/edge devices.
These points are related to each other by the available technologies. 
Approximately each decade has a technology becoming the mainstream in how computations are done. 
This list helps us to create the systematic view around different domains in Figure~\ref{figure:domains_X_Chart}.

\begin{itemize}[noitemsep,nolistsep]
    \item Trust Your Manufacturer in Confidential Computation: 
    \begin{itemize}[noitemsep,nolistsep]
      \item Local Confidential Computation.
    \end{itemize}
    \item The Use of Public Key Infrastructure in Remote Computing: 
    \begin{itemize}[noitemsep,nolistsep]
      \item Remote Representation of Confidential Computation. 
    \end{itemize}
    \item Old-school Cloud Computing as Confidential Computing: 
    \begin{itemize}[noitemsep,nolistsep]
      \item Partial Confidential Remote Computation in the Cloud. 
    \end{itemize}
    \item When Data Owners Claim Their Rights: 
    \begin{itemize}[noitemsep,nolistsep]
      \item Data-Confidentiality in Remote Computation.
    \end{itemize}
    \item Algorithms Matter: 
    \begin{itemize}[noitemsep,nolistsep]
      \item Fully General Confidential Remote Computation.
    \end{itemize}
\end{itemize}

From another perspective, \emph{\acrlong{crc}} can go back to remote procedure calls (RPC) in distributed computing.
Any functions executing in remote nodes need to comply with security requirements.
The following structure begins with local confidential computation (first stage), and the remote representation of confidential computation (second stage).

\subsubsection{First Stage: Trust Your Manufacturer in Confidential Computation }
Confidentiality requires at least two entities, one which protects an asset threatened by another. 
With regards to the earlier form of computation, it is possible to talk about confidential computation taking place in local machines. 
Individuals trust the manufacturers to build the correct computing device.
Yet, the manufacturers sell hardware and no direct networking takes place.
Individual computations take place locally and without external interactions in the early days.

\subsubsection{Second Stage: The Use of Public Key Infrastructure in Remote Computing}
\label{sec:PKIandCRC}
Around the 1970s\footnote{Online, Last Accessed 26 May 2022, \url{https://archive.nytimes.com/www.nytimes.com/library/cyber/week/122497encrypt.html}}\footnote{Online, Last Accessed 26 May 2022, \url{https://web.archive.org/web/19980507105259/http://www.cesg.gov.uk/ellisint.htm}}\footnote{Online, Last Accessed 26 May 2022, \url{https://web.archive.org/web/19980507105439/http://www.cesg.gov.uk/cnellis.htm}}, secure communication through certificate-based authentication  was initially developed/used by intelligence agencies\footnote{\eg~British Secret Service, or GCHQ, or considering some claims of NSA\@. Note added to clarify comments of one of the assessors of this work.}, but only became available to the general public by the 1990s\cite{singh2000code} through attempts of the public sector and researchers, for example, the invention of WWW by Tim Berners Lee (1991) and SSL/TLS by Taher Elgamal (1994). 
Certification Authorities (CAs) introduced a new trust model based on public key infrastructures. 
With the help of this infrastructure, increasingly complex trust models began to facilitate interactions in the digital world.
In those models, identity is provided with certificates.
We still have to trust the manufacturers to build the right hardware, but in addition, we also need to trust certification authorities to provide the advertised services.
While in the first stage local machine owners were only able to trust computations they performed themselves, now they can also validate and trust whether any given data, computation, application, or output stems from the expected entity.

\subsubsection{Third Stage: Old-school Cloud Computing as Confidential Computing}
Over time, entities with the ability of centralisation offered the first form of \emph{\acrlong{crc}}.
These services satisfy customers who need more computational power in order to outsource the computational jobs.
The set of hardware owned by those central authorities introduced a service we now know as cloud computing.
The problems in the threat modelling, (1) \emph{who protects what} and (2) \emph{protect from who} never became clear.
Two major problems appear after this stage: (1) Can the cloud providers trust their own infrastructure to do the right job? (2) Can customers trust the cloud provider's infrastructure to behave as promised?
In this setting, the customers must be able to trust to the remote computation, for example, a sorting, compression, or other services.
Private data is not necessarily involved yet.
Algorithm secrecy is not a concern yet either due to a lack of awareness of the possible value of data and algorithms.

\subsubsection{Fourth Stage: When Data Owners Claim Their Rights}

Soon after the widespread takeover of the cloud systems around 2010, with their existing problems, the value of data becomes more important.
The new evolving requirements bring new solutions together.
While in the previous stage of old school cloud computing, the data in use is not encrypted and therefore its usage carries high risks, at this stage, data owners require strong security guarantees for the data in use, besides data at rest and data in transit requirements.

\subsubsection{Fifth Stage: Algorithms Matter}
Recently, developers and companies asserting ownership on intellectual properties have become additional stakeholders in computations.
The intellectual property in question is the private algorithm and/or the business logic itself.
The algorithm security brings new challenges to private transactions, secret contracts, computer games, and many other domains.
Besides the secrecy of the algorithm itself, a public algorithm processing a piece of private data is an attack vector at risk of leaking private data through side-channel attacks.
It is not sufficient to keep algorithms private to mitigate the danger of side-channel attack, but data owners can benefit from secret algorithms depending on the decentralised trust model between hardware, data and algorithm owners.
Private algorithms must be concealed from the hardware owner to contribute to data security, because potential collusion of private algorithms and hardware owners poses a high risk to data owners. 
The case-specific threat model, however, must have a micro-optimisation depending on the available security features of the underlying hardware.

\section{Kernel and Hardware Assistance in Confidential Remote Computing}
\label{sec:kernelHardware}

\emph{\acrlong{crc}} has not been fully modelled yet, we listed the existing similar concepts in the Setion~\ref{sec:ambiguityofTrust}.
A way to achieve a comprehensive model can be by having the dimensions of \emph{assets} and \emph{demands} of mutually distrusting parties with conflicting interests.
Hardware owners, data owners and algorithm owners are distinguishable in a many-party computation.
These three parties are distinguished by the assets they bring to and the demands they expect of a computation.
In practical terms, any combination of these parties could be a single entity.
For example, a hardware owner and an algorithm owner may be represented by one company.
Hardware owners are concerned with maintaining their trustworthiness, maximising their revenue, and are honest-but-curious to learn anything about an algorithm (\ie~business or game logic).
The algorithm owners have interest in their logic (\ie~formulated computation) secret, and they are curious to learn about the data they process.
The data owners must protect the privacy and secrecy of the data set they manage, while aiming to maximise their revenue at the same time.
Both data and algorithm owners are potentially interested in the outputs of computation.

In considering the security of distributed computing, we may trace a development path from the remote procedure call (RPC) forward.  
The earliest problem to be addressed was managing access control for the remote computational resource, followed by technologies for the security of communications to and from the remote platform.  
In the simple case, attestation adds further assurance to the initiating party about the service offered on the remote system.  
Moreover, attested service isolation helps to provide assurance for credentials passed to the remote service so that that service can read or write from other resources.

Much of our work has concentrated on a generalisation of multiple parties wishing to collaborate on a task, despite being mutually-distrusting (and distrusting the remote platform also).   
Attestation provides a mechanism to enable them to gain trust in the remote platform, but raises many challenges for the management of general-purpose platforms.


\begin{figure}[htp!]
  \centering
  \includegraphics[width = 1\textwidth]{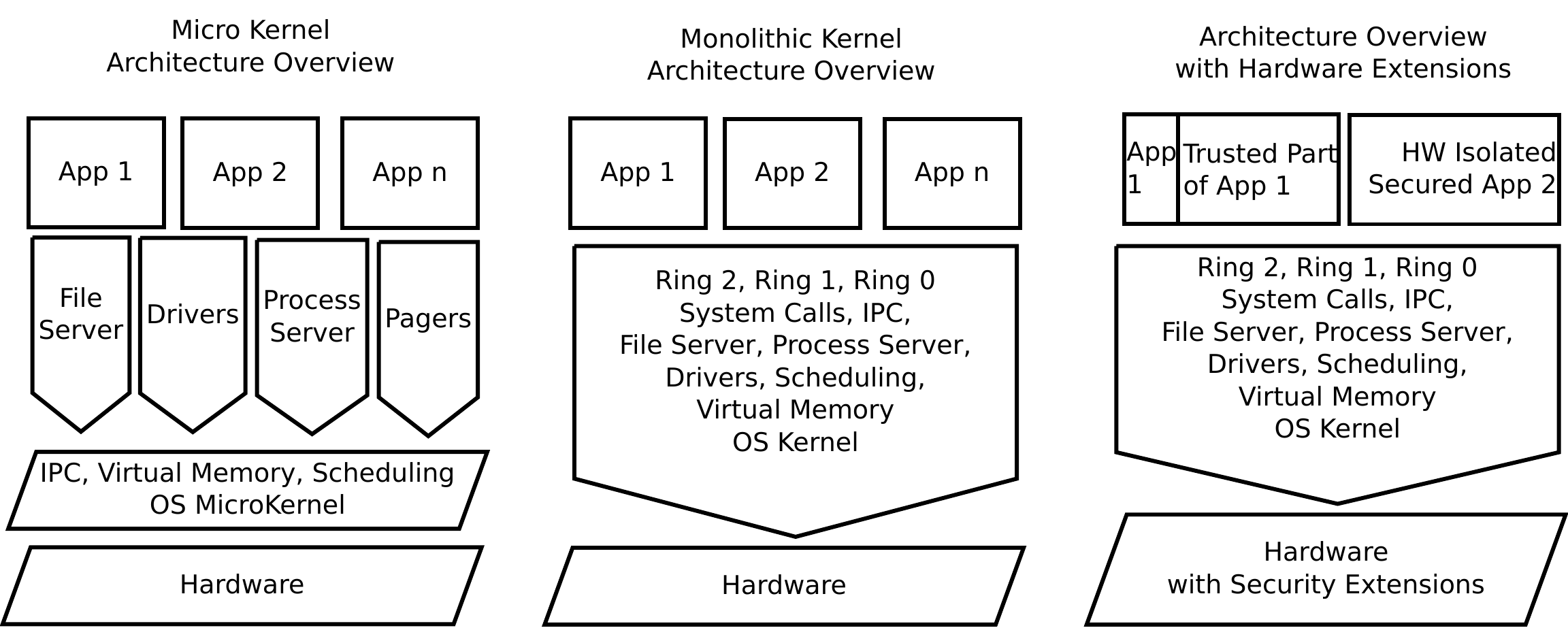}
  \caption{Overview of system architectures for security and performance.
  Monolithic kernels offer better performance, and, their security limitations can be mitigated with hardware extensions.}\label{fig:kernels}
\end{figure}

In this \chapterPaper{chapter}{paper}, we explore the most general case, then, where the data \emph{and} the computation are both secrets.  
Clearly, in this case, binary attestation may be sufficient to assure all parties that they are interacting with the same code entity, but crucial questions about the behaviour of the computation can only be addressed through some form of semantic remote attestation.

Our objective in this \chapterPaper{chapter}{paper}~is to explore these dimensions and provide an overview of potential strategies for trustworthy remote computation, leading towards a fully-general \emph{\acrlong{crc}} framework, always having regard for the concerns of \acrshort{tcb} minimisation. 
We extend the framework to consider the notions of privacy protection explored elsewhere in the project as examples of the semantic properties to be attested.

\subsection{What benefits can Hardware-based Root of Trust offer?}

Basing the root of trust in hardware may offer stronger security guarantees than software-based mechanisms for attestation (local or remote) during a computation.

The core of security mechanisms in systems relies on components specified in the Root of Trust (RoT).
Sub-definitions of RoT such as \acrlong{rtr}, \acrlong{rtm} or \acrlong{rts} offer increased security  with their independence and low size.
Establishing trust in a system may require a piece of evidence~(\eg~hash, checksum) representing the state of the memory.

\subsubsection{Is Software-based Root of Trust Insufficient?}

In embedded systems or in computer peripherals, the software-based attestation mechanisms~\cite{Seshadri2004,Li2010} may generate time-based reliable reports about the trustworthiness of the system.
General-purpose computers are more complex and have larger attack surfaces, and software-based attestation mechanisms fail~\cite{Jakobsson2011,Castelluccia2009} to represent the state of the system. 
Software-based attestation mechanisms cannot provide secure reports in rich execution environments if the attacker has physical access to the system.
Alternatively, a software-based trust may fail if any exploit gives attackers equal privileges to attempt an attack on a lower level (towards hardware) than the software-based attestation mechanism runs at.

Hardware-based attestation mechanisms often run on the lowest level of systems and utilise private keys embedded inside the chip to generate trusted pieces of evidence representing the memory state. 
In our model, we focus on remote attestation mechanisms based on secure processors described in Section~\ref{sec:att_domain}.

\subsection{Failure or Success of the Micro-kernels?}

The second architectural question for \emph{\acrlong{crc}} concerns the kernel and privilege structures responsible for performance and security.

Micro-kernel architectures provide software isolation for kernel functionalities, shown in Figure~\ref{fig:kernels}.
In the execution phase, servers do not share resources or pointers and do not have direct access to the address space of other servers. 
Even in a traditional or layered kernel architecture, address spaces are isolated from each other with privilege levels; if one server passes any address space to another kernel module, changes must be tracked and controlled.
The maintenance cost of memory tracking increases the performance overheads in micro-kernels.
Even with an acceptable performance overhead, they require the utilisation of distributed programming models to share and maintain the resources between several modules. 
The distributed algorithms may not be sufficient to make the system available; if a kernel module fails, the system may fail too because of dependent resources, as occurs in distributed systems.

Besides the notion of availability, security is also characterised by the notion of non-compositionality~\cite{Cremers2006}. 
Having individual secure kernel modules in the micro-kernel architecture approach does not prove the overall security of the kernel.
The security properties of smaller modules may not make the system containing it secure by themselves.
Micro-kernels were developed later than the monolithic kernel approach.
Current computer systems are mostly using monolithic kernel architectures, and deployment of systems based on micro-kernel architectures requires significant development effort.

The Information Flow Control (IFC) kernels~\cite{Zeldovich2007,Pasquier2016} can establish a trusted state for applications running on top of untrustworthy code.
Their trust assumptions include hardware integrity and physical security against tampering.

The wimp-giant kernel architecture model~\cite{10.1007/978-3-319-12400-1_11} suggests an abstract isolation mechanism assuming to provide accurate and complete adversary definitions.
In this approach, micro-hypervisors require a verifiable boot.
Otherwise wimpy hypervisors and wimpy kernels cannot bring any security guarantee for wimp apps.
A hardware-based remote attestation mechanism may satisfy the verifiable boot requirement of a wimpy micro-hypervisor.
Nevertheless, this approach does not provide confidentiality and integrity properties for wimp apps.
A malicious giant may read the contents of wimpy apps as they please.
A giant may compromise the security of wimpy apps with a TOCTTOU attack~\cite{Weichbrodt2016} or an IAGO attack~\cite{Checkoway2013}.
Due to incomplete adversary definitions, the dancing wimp-giant kernel architecture model is insufficient for establishing trust in the presence of the untrusted environment.

A micro-kernel alone may not be able to offer the security guarantees of hardware-assisted systems.
But micro-kernels may be more suitable for formal verification than monolithic kernels.

The formally verified micro-kernels can offer isolation and confidentiality guarantees similar to enclave-based (hardware-assisted) systems.
Besides, micro-kernels can be integrated with hardware extensions in a system for better security.
For example, the seL4~\cite{klein2009sel4} formally verified micro-kernel can be used with RISC-V~\cite{Asanovic:EECS-2016-17} systems with the Keystone Enclaves~\cite{lee2020keystone}.
Either the underlying system managing the enclaves can utilise the seL4, or the seL4 instance can be placed inside the enclaves for kernel support.
However, one of the difficulties with formal verification of the micro-kernels remains unsolved, major updates or changes in the kernel may invalidate its proofs.
Frequent modifications to proofs limit the implementation of sophisticated functionalities in micro-kernels and thereby their wide deployment in general purpose computers.

\subsection{Monolithic kernels are faster but offer no better security than micro-kernels}

The X86 monolithic kernel architecture includes four privilege rings: operating system kernel level at ring 0, rest of the operating system at ring 1, device drivers at ring 2 and applications at ring 3.
Most of today's systems use only ring 0 including the entire kernel, OS, and drivers, and ring 3 for user applications.
Ignoring ring 1 and ring 2 brings performance benefits, as interrupts cause switch overheads between the rings.
Since most of the existing operating systems in use are already designed to use only ring 0, it is impractical to re-implement all their software stack.
The lack of intermediate privilege rings increases the security risk.
If any user application can jump into ring 0, it takes full control of the operating system.
If a malicious application hooks the system before a security mechanism running on ring 0 can do so, this means the security mechanism cannot detect the malicious application.
This is one of the fundamental problems most anti-virus software products suffer from. 
A ring-3-malware may easily take control of the kernel in ring 0, and fully evade its anti-virus software.

\subsection{Improving security of monolithic kernels with protected module architectures}

Protected module architectures (\eg~Intel's recent iteration) utilise the ring -3 (minus three) and increase the security of user-level applications with the ability to sandbox a memory region from the rest of the system.
The microcode of the processor handles the isolation mechanisms of the user-level applications running on ring 3. 
This trend brings stronger security guarantees to existing systems and its development models allow existing applications to run with no modification or with small changes.
The applications deployed on secure processors can also benefit from the memory encryption engine, remote attestation, confidentiality and integrity features.

\section{Practical Implementations of Enclaves}\label{sec:enclavesInPractice}
Enclaves are the protected memory regions containing an application code.
The protected module architectures provide the security features for these enclaves.

\begin{figure}[htp]
\centering
  \includegraphics[width=0.6\linewidth]{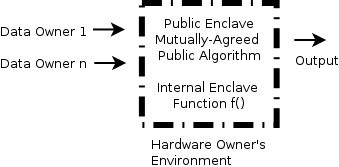}
  \caption{
  Many Data Owners join into a mutually agreed multi-party computation.
  An enclave can act as a trusted proxy for computations.}\label{fig:multipleData}
\end{figure}

\begin{figure}
  \centering
  \includegraphics[width=0.6\linewidth]{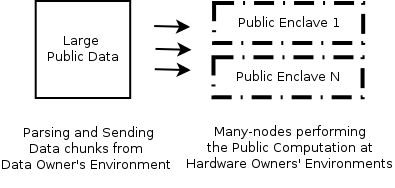}
  \caption{Multiple Hardware Owners contribute to grid Computation. 
  Enclaves can prove the integrity of a completed job.}\label{fig:gridComputing}
\end{figure}

We describe two uses of enclaves: first grid computing, and second multi-party computing.
In the grid computing shown in Figure~\ref{fig:gridComputing}, multiple hardware owners using enclaves can process a big data silo and prove that they did their job. 
In the multi-party computation shown in Figure~\ref{fig:multipleData}, multiple data owners may jointly perform computations over their inputs.

\subsection{Multi-Enclave Grid Computing}
A long lasting question in grid computing can be answered by looking at the use of enclaves: how can we make sure that a given software processing a significant amount of non-sensitive data (\eg~satellite, climate data) is running on non-tampered computers of end-users contributing to computation?
Developers can send the compiled enclave binary to the nodes (who make revenue by running the software on their hardware), and developers can attest integrity of the enclaves with binary-based remote attestation mechanisms.
If an enclave algorithm is public, and if the processed data is public, enclaves can provide great integrity guarantees for computations.

\subsection{Trust in Grid Computing, Edge Computing, Fog Computing}

In grid computing, the nodes are assigned a task containing a computation and must prove the integrity of the computation.
During the lifecycle of the taken job, the computing farm must satisfy the integrity requirement.
Confidentiality is also required, as the data is sent to the computing farm.
On the other hand, in edge computing, edge devices' own data is processed in hardware before sending it to the remote or cloud environment.
Later on, the remote environment can still trust the validity of received data and integrity of computation.
For example, in the IoT domain, computational correctness is in the interest of the IoT device, the node itself or the system.
Taking as an example a smart car system with a fog computing model, the system must display the correct behaviour while sending data to the remote entities. 
This is required because otherwise the computations can be altered, potentially leading to a loss of revenue.
All these systems can benefit from trusted hardware primitives and satisfy their integrity requirements. 

\subsection{Multiple Data Owners using Enclaves}

Similarly, multiple \emph{Data Owners} can join a computation utilising the integrity guarantees of the enclaves.
If the data is not privacy-sensitive or confidential, the \emph{Data Owners} can retrieve cryptographic evidence about the data sets used in a mutually agreed on computation using a publicly visible algorithm.
This way, each participant may trust the performed operations.
However, if the data is privacy-sensitive, the question would be whose physical environment will be in use for the computation?
The physical environment of execution shows the entity who controls the physical hardware and the system used to define the threat model.

To process confidential data on public algorithms deployed in public enclaves, the enclave must be programmed carefully against non-trivial side-channel attacks (\eg~page-level memory attacks) depending on local or remote \emph{Physical Environment} of execution.
Enclaves can provide integrity, but only limited confidentiality for a computation.
Developers may provide additional confidentiality with secure programming.

\subsection{Various Aims of Hardware Owners for Enclave Software}

\emph{Hardware Owners} differ by the attestation mechanisms they use.
For example, the precise attestation information about a system or software is helpful for a company to identify the system version. 
But, on the other hand, it can tell adversaries the system version as well, thus informing them of existing vulnerabilities.
Furthermore, home users require less precision of their attestation reports, although they may still need higher levels of precision to prove what system or software they are running.
In the short Table~\ref{tab:hardware_owners_differ}, we highlight the differences of hardware owners in attestation aspects.
A key difference is that, as opposed to home users, enterprises or military organisations often keep their attestation evidence and communications internal within their respective intranets, maximising control over their own systems.
In contrast, home users interact frequently and obliviously with potential threats.
This raises privacy concerns which can be addressed by requiring the attestation mechanisms to remove the identification information of users.

\begin{table}[hbpt]
    \centering
    \resizebox{1\textwidth}{!}{%
    \begin{tabular}{ l | c | c }
       Type of Hardware Owner  & Big Enterprise / Military & Home User   \\ \hline
       In Attestation &  Prioritises Precise Information & Anonymous Hardware IDs  \\ \hline  
       Privacy Aspect &  No Privacy Concerns & Needs Privacy  \\ \hline
       Interaction & Internal & External \\ \hline
    \end{tabular}
    }
    \caption{Big enterprises and home users require different attestation evidences.}\label{tab:hardware_owners_differ}
\end{table}

\subsection{Confidential Remote Computing in the Real World}

We give a few exemplary systems to map \emph{\acrlong{crc}} trade-offs in the real world.
First, a smart grid system, the measurements can be collected every \emph{30 minutes}, potentially for a town of \emph{20.000 people}, for the privacy-preserving operations for billing, monitoring, and demand-response operations.
Second, as another example, we may take a government election. 
During a voting day, the core task for a city of \emph{10 million citizens} is to complete the election in \emph{8 hours}, including the casting, verifying, and counting operations.
These computations eventually take place in the remote computers where all participating entities can be satisfied with the level of decentralisation and the evidence of transparency.
Apart from this core task (\eg~vote-casting, verifying, counting), a system can also be initially defined by the level of decentralisation or transparency users (\eg~voters) require.
In a complex system of IoT deployment, \emph{Confidential Remote Computation} can be defined with hundreds of computing devices running algorithms provided by different entities, and processing data of multiple users.
In an industrial IoT deployment, the number of hardware devices can go up to millions.
All these continuous computations (also called long-term jobs) represent one single \emph{\acrlong{crc}} system.
Long-term jobs may contain private optimisation and minimisation algorithms for manufacturing complex geometries~\cite{zschippang2019face,zschippang2020face}.
The input parameters of desired geometries, algorithm itself, output models and machine behaviour data must remain secret.

The COVID-19 track and trace systems can be built in the \acrshort{crc} model. 
People provide their location, time and phone number per shop in exchange for a notification service, and are promised deletion of their sensitive data after its expiry.
Despite these purported assurances, however, there is little or no effective privacy protection for users' behavioural data.
The data collection points (\eg~restaurants) and the data processing points (government health authority) provide no reliable evidence of secure data processing, deletion and refraining from abuse (using it for any other purposes, even anonymised).
The computation time can be seen as \emph{2 weeks per check-in} from an individual's perspective.
The expected behaviour of the system is to scan the collected data backwards for each data collection point and to notify relevant individuals whenever a new case or patient is detected.
Although ideally this process could be set up with extensive security and privacy services at the cost of becoming more computationally expensive, in practice it is built in a straight-forward manner, with its plain-text data processing taking place in a central server whose hardware owners can observe all documented user activities unhindered.

\clearpage

\section{Co-Evolution of Requirements and Solutions through Five Entities}\label{sec:xchart}

In order to model \emph{\acrlong{crc}}, we need to go through three domains as follows: \textbf{the hardware domain} as a root of trust, \textbf{the development domain} for software composition and \textbf{the attestation domain} for integrity representation.
We systematise these domains by utilising five distinct parties directly or indirectly involved in the computation in the following manner:
\begin{enumerate}[noitemsep,nolistsep]
    \item \textbf{Manufacturer:} Local Confidential Computation
    \item \textbf{Certification Authority:} Remote Presentation of Confidential Computation
    \item \textbf{Hardware Owner:} Partial Confidential Remote Computation in the Cloud
    \item \textbf{Data Owner:} Data-Confidentiality in Remote Computation
    \item \textbf{Algorithm Owner:} Fully General Confidential Remote Computation
\end{enumerate}
Each of these participants introduces new requirements and new trust models into a computation.
We use each participant's respective requirements in chronological order for our model. 
This classification of computation types offers new concepts and solutions for the three domains outlined above.

\begin{figure}[!ht]
  \centering
  \resizebox{1\textwidth}{!}{%
  \begin{tikzpicture}[>=stealth',join=bevel,auto,on grid,decoration={markings,
    mark=at position .05 with \arrow{>}}]
    \large
    \coordinate (attestationNode) at (45:6cm);
    \coordinate (developmentNode) at (135:6cm);
    \coordinate (hardwareNode) at (225:6cm);
    \coordinate (participantNode) at (315:6cm);

    \coordinate (originNode) at (0:0cm);

    \node [above=1em] at (developmentNode) {\textbf{2. Development Domain}};
    \node [below=1em] at (hardwareNode) {\textbf{1. Hardware Domain}};
    \node [above=1em] at (attestationNode) {\textbf{3. Attestation Domain}};
    \node [below=1em, text width=5.4cm] at (participantNode) {\textbf{Requirements Evolve with each Entity}};

    \draw[-, very thick, arrows={<-[length=10]}] (developmentNode.south) -- (0,0)
    node[left,pos=0.1]{Secure Loader}
    node[left,pos=0.25]{TCB Size}
    node[left,pos=0.40]{Interfaced Applications}
    node[left,pos=0.55]{Unmodified Applications}
    node[left,pos=0.70]{OS Libraries}
    node[left,pos=0.85]{Licensing} ;

    \draw[-, very thick, arrows={<-[length=10]}](hardwareNode.south) -- (0,0)
    node[left,pos=0.05, anchor=east] {TCB Update}
    node[left,pos=0.15, anchor=east]{Private RAM}
    node[left,pos=0.25, anchor=east]{Initialise First}
    node[left,pos=0.35, anchor=east]{Multiple-Isolation}
    node[left,pos=0.45, anchor=east]{Mobility}
    node[left,pos=0.55, anchor=east]{Native CPU}
    node[left,pos=0.65, anchor=east]{Arbitrary Code}
    node[left,pos=0.75, anchor=east]{Quoting Speed}
    node[left,pos=0.85, anchor=east]{Maturity}
    node[left,pos=0.95, anchor=east]{Cost Efficiency};

    \draw[-, very thick, arrows={<-[length=10]}] (attestationNode.south) -- (0,0)
    node[right,pos=0.1]{Frequent Updates}
    node[right,pos=0.25]{Secrecy}
    node[right,pos=0.4]{Accuracy}
    node[right,pos=0.55]{Request Overhead}
    node[right,pos=0.7]{Forward-Reliance}
    node[right,pos=0.85]{External Calls} ;

    \draw[-, very thick, arrows={<-[length=10]}] (participantNode.south) -- (0,0)
    node[left,pos=0.1, ,anchor=west]{Algorithm Owner }
    node[right,pos=0.3, ,anchor=west]{Data Owner}
    node[right,pos=0.5, ,anchor=west]{Hardware Owner}
    node[right,pos=0.7, ,anchor=west]{Certificate Authority}
    node[right,pos=0.9, ,anchor=west]{Manufacturer};

    \draw[fill] (barycentric cs:developmentNode=0.9,originNode=0.1) circle (2pt);
    \draw[fill] (barycentric cs:developmentNode=0.75,originNode=0.25) circle (2pt);
    \draw[fill] (barycentric cs:developmentNode=0.6,originNode=0.40) circle (2pt);
    \draw[fill] (barycentric cs:developmentNode=0.45,originNode=0.55) circle (2pt);
    \draw[fill] (barycentric cs:developmentNode=0.3,originNode=0.7) circle (2pt);
    \draw[fill] (barycentric cs:developmentNode=0.15,originNode=0.85) circle (2pt);

    \draw[fill] (barycentric cs:hardwareNode=0.93,originNode=0.07) circle (2pt);
    \draw[fill] (barycentric cs:hardwareNode=0.85,originNode=0.15) circle (2pt);
    \draw[fill] (barycentric cs:hardwareNode=0.75,originNode=0.25) circle (2pt);
    \draw[fill] (barycentric cs:hardwareNode=0.65,originNode=0.35) circle (2pt);
    \draw[fill] (barycentric cs:hardwareNode=0.55,originNode=0.45) circle (2pt);
    \draw[fill] (barycentric cs:hardwareNode=0.45,originNode=0.55) circle (2pt);
    \draw[fill] (barycentric cs:hardwareNode=0.35,originNode=0.65) circle (2pt);
    \draw[fill] (barycentric cs:hardwareNode=0.25,originNode=0.75) circle (2pt);
    \draw[fill] (barycentric cs:hardwareNode=0.15,originNode=0.85) circle (2pt);
    \draw[fill] (barycentric cs:hardwareNode=0.05,originNode=0.95) circle (2pt);

    \draw[fill] (barycentric cs:attestationNode=0.9,originNode=0.1) circle (2pt);
    \draw[fill] (barycentric cs:attestationNode=0.75,originNode=0.25) circle (2pt);
    \draw[fill] (barycentric cs:attestationNode=0.6,originNode=0.4) circle (2pt);
    \draw[fill] (barycentric cs:attestationNode=0.45,originNode=0.55) circle (2pt);
    \draw[fill] (barycentric cs:attestationNode=0.3,originNode=0.7) circle (2pt);
    \draw[fill] (barycentric cs:attestationNode=0.15,originNode=0.85) circle (2pt);

    \draw[fill] (barycentric cs:participantNode=0.9,originNode=0.1) circle (4pt);
    \draw[fill] (barycentric cs:participantNode=0.7,originNode=0.3) circle (4pt);
    \draw[fill] (barycentric cs:participantNode=0.5,originNode=0.5) circle (4pt);
    \draw[fill] (barycentric cs:participantNode=0.3,originNode=0.7) circle (4pt);
    \draw[fill] (barycentric cs:participantNode=0.1,originNode=0.9) circle (4pt);


  \draw[domain=0:25,variable=\t,samples=440,smooth,arrows={->[length=10]}]
        plot({(-\t-0.8) r}:{\t/5+0.9*\t/(0.1+\t)});

  \end{tikzpicture}
  }
  \caption{Confidential Remote Computing X-chart systematised around the increasing demands of five participants.
  Evolving concepts and solutions are classified under three domains.
  These domains show the orthogonal research directions of the field.}
\label{figure:domains_X_Chart}
\end{figure}
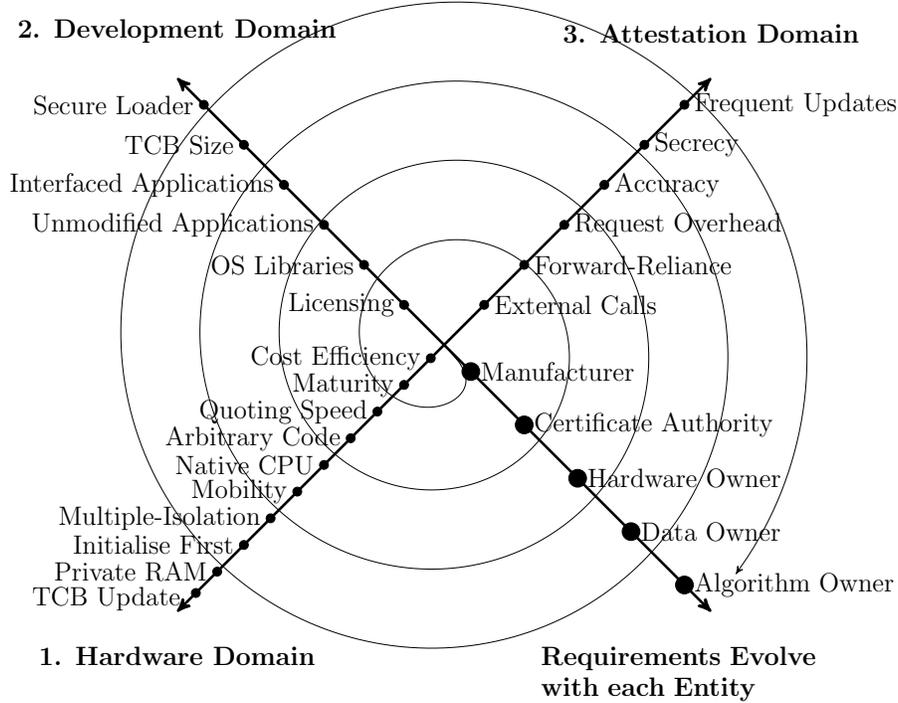

\subsection{Three Domains of Confidential Remote Computing}

These three domains are aligned with the dimensions of \emph{\acrlong{crc}}: hardware features, programming models and attestation mechanisms. 
Figure~\ref{figure:domains_X_Chart} displays the three domains of \emph{\acrlong{crc}}, shows how we distinguish the methods, and provides an extendable picture for future methods under these domains.
The elements of each domain are listed in order of availability represented by points along the axes of the figure.
Elements arranged closer to the centre have already been adopted at large by the industry at the time of writing.
Elements closer to the outer circle may not be available in the market yet and/or have been only recently introduced, but are relevant nevertheless as we shall demonstrate shortly.
Each circle completed through the dimensions represents a time period during which the included elements become available on the market as part of technology products.

Our X-chart in Figure~\ref{figure:domains_X_Chart} provides a systematised image of the domains of \emph{\acrlong{crc}} comprising the following nodes:
\begin{itemize}[noitemsep,nolistsep]
\item Ten metrics for hardware technologies
\item Five features of development techniques
\item Six benefits of attestation mechanisms
\item Five distinct roles of participating entities~(with conflicting interests)

\end{itemize}

The following four sections elaborate on the four dimensions of this chart.
In Section~\ref{sec:hw_domain}, we describe the features of the \textbf{hardware domain}.
This domain includes the practical hardware functionalities for \emph{\acrlong{crc}}.
Each element described in this domain offers different benefits for a target system.
They may improve performance, integrity or confidentiality of the system, reduce the cost of development, or give stronger cryptographic evidence of the trustworthiness of the system.
Each feature is implemented in at least one hardware solution on the market, but no single hardware solution comprising all the features described has been introduced on the market.
An intended \emph{\acrlong{crc}} system may use one or more hardware products in combination to address stronger adversary models.
We describe how an ideal system could utilise each hardware functionality and identify the ideal trusted hardware to meet the demands of \emph{\acrlong{crc}}.

We describe the \textbf{development domain} in Section~\ref{sec:dev_domain}.
For the development of \emph{\acrlong{crc}} applications, there are a few different possible programming models available in the field of trusted computing.
Each method may offer different benefits such as providing a smaller trusted computing base, license-free development or a language defining the system's interaction with the outside world.
Evaluating these benefits, we identify the ideal programming model for a \emph{\acrlong{crc}} system.

Section~\ref{sec:att_domain} addresses the \textbf{attestation domain}.
We describe the benefits of different attestation mechanisms for \emph{\acrlong{crc}}.
The various available mechanisms provide evidence of different aspects of a computation, different accuracy levels for trustworthiness, cause different overheads, handle different communication flows, and give different guarantees for attestation.
We explain why a single attestation mechanism is not sufficient for \emph{\acrlong{crc}}.
We also discuss the potential attestation domain aspects of a hypothetical ideal \emph{\acrlong{crc}} application.

We present \textbf{the conflicting interest of participants} in Section~\ref{sec:prt_domain}.
The \emph{\acrlong{crc}} model involves several different participants, each holding different assets to protect and different demands to maximise.
We distinguish the participants by their role in the \emph{\acrlong{crc}} systems.
There might be tens of thousands of parties of a certain type of participant group involved with any one \emph{\acrlong{crc}} system at higher scale.
We identify three major participant types (\ie~hardware, data, algorithm owners), aside the minor less numerous participants (\ie~Certification Authorities and hardware manufacturers).
We explain the characteristics of the major participant types, and introduce potential scenarios how each participant may maximise their revenue while still protecting their assets. 
We also explain why certain configuration models will lead to an inequitable arrangement.

\subsection{Hardware Features}\label{sec:hw_domain}

Trusted hardware provides a more secure environment for execution compared to standard hardware.
However, any single trusted hardware alone is not sufficient for an ideal \emph{\acrlong{crc}} application.
We compare three leading trusted hardware products by the features we identified for \emph{\acrlong{crc}} applications.
Table~\ref{compareHW} lists hardware domain elements useful for \emph{\acrlong{crc}} and compares three hardware technologies widely available on the market: Intel \acrshort{sgx}~\cite{SGXrefProgramming}, ARM Trustzone~\cite{Alves2004} and the \acrshort{tcg} \acrshort{tpm}~\cite{tpmSpec} as three examples of available hardware.
We pick Intel \acrshort{sgx} for its availability on desktop computers, ARM Trustzone for its mobility on the IoT and smartphone market, and \acrshort{tpm} because it is an independent external chip performing trustworthy operations.

For sake of brevity, we omit other secure processors such as XOM~\cite{lie2000architectural}, Aegis~\cite{suh2007aegis}, Bastion~\cite{champagne2010scalable}, Ascend~\cite{fletcher2012secure}, Phantom~\cite{maas2013phantom}, Sanctum~\cite{costan2016sanctum} in this \chapterPaper{chapter}{paper}.
Future work may evaluate these processors along the same hardware dimension we described.
We also omit the technologies TDX~\cite{lal2019technologies,ghosh2019systems}, Morello~\cite{arm2020Morello}, AMD SEV (ES and SNP)~\cite{sev2020strengthening} and IBM PEF~\cite{ibm2020PEF}, which run towards similar aims as \acrshort{sgx} and Trustzone.
A detailed analysis of these trusted hardware technologies may be explored elsewhere.

\textbf{Multiple-Isolation:} A secure application must be isolated from any untrusted component.
As an external chip, \acrshort{tpm} does not provide any isolation for the computation, but it provides secure storage for the measurements (\acrlong{rts}) of a computation~\cite{tpmSpec}.
\acrshort{tpm} itself does not isolate the system from applications, but \acrshort{drtm} mechanisms can be used to have software-based isolation mechanisms for applications.
ARM Trustzone~\cite{Alves2004}-based processors provide strong isolation for a single secure world placed in a secure memory area which is different from main memory~\cite{ARM2009}, but the hardware itself does not provide multiple memory regions for isolation. 
There are \acrshort{tee}~\cite{Asokan2013,teeSpec} software solutions such as Trustonic~\cite{trustonicDocs} for isolated multiple execution environments~\cite{trustonicKinibiTechnical,Alvarado2008,Apvrille2011}. 
Even so, software-based solutions may not~\cite{Atamli-Reineh2016,zeroTrustonicIssues} be sufficiently protective against powerful adversaries.
\acrshort{sgx} as implemented by Intel provides a multiple-isolation mechanism in microcode for its enclaves~\cite{SGXrefProgramming}; this provides multiple small regions for computation isolated from each other. 
However, Intel's solution does not provide a fully private memory; all enclaves sit on the main memory.

\textbf{Private RAM:} A secure system must have its own secure memory area, fully differentiated from any of the untrusted components.
\acrshort{tpm} has its own storage for measurements~\cite{tpmSpec}, but it does not offer any benefit for a computation loaded and executed in the main system.
\acrshort{tpm} can help to ensure software based memory encryption mechanisms are in place, but application's memory still sits in the main memory.
ARM Trustzone has the best solution among trusted hardware solutions on the market. 
The secure world sits in a different memory area (depending on manufacturers' setting) than the main memory~\cite{ARM2009,ARM2016}. 
The operating system and any untrusted component (normal world) of an ARM Trustzone enabled device has no control over the protected memory of the secure world~\cite{ARM2009,Alves2004}.
Intel's \acrshort{sgx} has no private memory solution. 
All of the enclave memory uses pages within the main memory~\cite{SGXenclaveWritingGuide}. 
This leaves access patterns of memory open to inspection by potential attackers.
By design, Intel's processors also have various shared resources which is a crucial shortcoming for security-critical applications.

\textbf{Mobility:} A secure IoT solution needs good mobility for deployment in the real world.
\acrshort{tpm} is an external chip accommodated on the system board~\cite{tpmSpec}; \acrshort{tpm} is widely included in most modern hardware.
ARM Trustzone-equipped devices provide mobility and excellent security features in the mobile environment~\cite{Alves2004}. 
ARM processors are low-energy consuming and widely used in most of the currently available mobile devices.
Intel's \acrshort{sgx} offers almost no mobility for IoT yet, as it is only deployed in high energy-consuming processors which require more power than its competitors, and Intel machines are larger by size.

\textbf{Fast Quoting:} In order to  attest a remote setting, hardware must generate cryptographic evidence about the trustworthiness of a system.
\acrshort{tpm}-based systems have quoting speed of 731ms~\cite{paverdThesis}. This quoting speed may fulfil the requirements of many party applications, however, at larger scale, it may not suffice as shown in recent studies~\cite{Kucuk2016}.
ARM Trustzone does not have any quoting or attestation mechanism for remote systems.
Developers may implement a custom software-based attestation mechanism, and ARM Trustzone can manage a secure boot and measurement for software attestation~\cite{ARM2016}.
However, a custom attestation implementation would not provide security guarantees of comparable strength to those enhanced within the hardware.
Intel's \acrshort{sgx} has the fastest quoting speed so far available on the market at around 20-30ms~\cite{Kucuk2016} speed per quote\footnote{even faster-quoting speed by the time of writing this document, thanks to recent fixes in Intel SDK}, which currently makes it the best candidate for deployment of \emph{\acrlong{crc}} applications.

\begin{table}
  \centering
  \resizebox{0.5\columnwidth}{!}{%
  \begin{tabular}{l c c c c}
    \toprule
    Ideal Feature/HW      &    (1)  &  (2)     &   (3)      &   Future      \\
    \midrule
    Multiple-Isolation   &   Y     &     N            &    N        &    Y       \\ \hline
    Private RAM          &   N     &     Y            &    N        &    Y       \\ \hline
    Mobility             &   N     &     Y            &    Y        &    Y      \\ \hline
    Fast Quoting         &   Y     &      N           &    N        &    Y       \\ \hline
    Initialize First     &   N     &     Y            &    Y        &    Y     \\ \hline
    Native CPU           &   Y    &      Y            &    N        &     Y     \\ \hline
    TCB Update           &   Y    &       N           &     N       &     Y      \\ \hline
    Arbitrary Code       &   Y     &      Y           &     N       &     Y      \\ \hline
    Secure Loader        &   N     &      N          &      N       &     Y     \\ \hline
    Maturity             &   N     &      Y          &      Y       &     Y     \\ \hline
    Cost Efficiency                &   N    &       N          &      Y       &     Y     \\
  \bottomrule
\end{tabular}
}
\caption{Comparing three trusted hardware models: Intel SGX (1) in-processor TEE, ARM Trustzone (2) mobile TEE, and TPM (3) external trusted hardware from TCG.}\label{compareHW}
\end{table}

\textbf{Initialise First:} 
Initialisation of the security module supporting applications should take place before the execution of any untrusted component in order to prevent any hook being placed against trusted components. 
Alternatively, the initialisation of the trusted component should not be dependent on untrusted components.
\acrshort{tpm} does not aim to provide direct assistance for the initialisation process of secure computations, but it runs externally and independently of the system; its features themselves are available for computations.
ARM Trustzone offers the best solution so far presented on the market, as its hardware initialises the \emph{Secure World} before the \emph{Normal World}~\cite{ARM2016}.
Moreover, its \emph{Secure World} application can~\cite{ARM2009} set a timer to trigger itself through an interrupt at any point in time without relying on a dependency on the \emph{Normal World}.
This feature is one of the most significant advantages of ARM hardware in deploying \emph{\acrlong{crc}} applications on Trustzone-enabled devices.
Intel's \acrshort{sgx} has no independence from the untrusted world.
The operating system has to trigger and allocate memory for the enclaves~\cite{SGXdevGuide21}.
Intel's \acrshort{sgx} relies on their attestation mechanism to verify the integrity of the enclave later on.
However, the OS and untrusted components still have full control over enclaves.

\textbf{Native CPU:} A secure application may require high processing power for its computations.
\acrshort{tpm} is not a processor and does not provide any computational resources for the execution environment.
ARM Trustzone processors feature relatively good CPU power that a \emph{Secure World} can utilise in mobile systems.
Intel's \acrshort{sgx} offers the highest CPU power for enclaves so far on the market; it is also available for server processors.

\textbf{TCB Update:} The \acrlong{tcb} of a system may need updates from time to time due to emerging vulnerabilities or newer functionalities.
\acrshort{tpm} may offer a \acrshort{tcb} update with append-only storage of hashes assisting with the representation of the updated state of a certain system or programme.
It offers fundamental RoT primitives to identify the system's state or the state of arbitrary computations.
System designers must define and load the software of the \emph{Secure World} in advance, and its \acrshort{tcb} update may be limited to patching the \emph{Secure World}.
Custom solutions can be implemented by the TEE developers.
Intel's \acrshort{sgx} provides \acrshort{tcb} update mechanisms for both the \acrshort{sgx} hardware and the enclaves~\cite{SGXupdateTCB}. 
\acrshort{tcb} updates help patch the CPU Firmware (microcode) to fix the hardware bugs.
Enclaves derive benefits from secure updates. 
For example, CPU can refuse executing outdated enclave software.

\textbf{Arbitrary Code:} A \emph{\acrlong{crc}} application may require arbitrary code execution on trusted hardware.
Computations take place outside of \acrshort{tpm} in the system. 
\acrshort{tpm} does not handle the computations itself, but \acrshort{tpm}-based solutions may assist applications in general-purpose computers.
ARM Trustzone-based systems require developers to place the application code in the \emph{Secure World} in advance, and the participants may not be able to load any arbitrary code remotely later on without third-party solutions.
Intel's \acrshort{sgx} allows~\cite{SGXdevGuide21} users to load any arbitrary code into the enclaves.

\begin{figure*}[htbp]
  \centering
  \includegraphics[width =  1\textwidth]{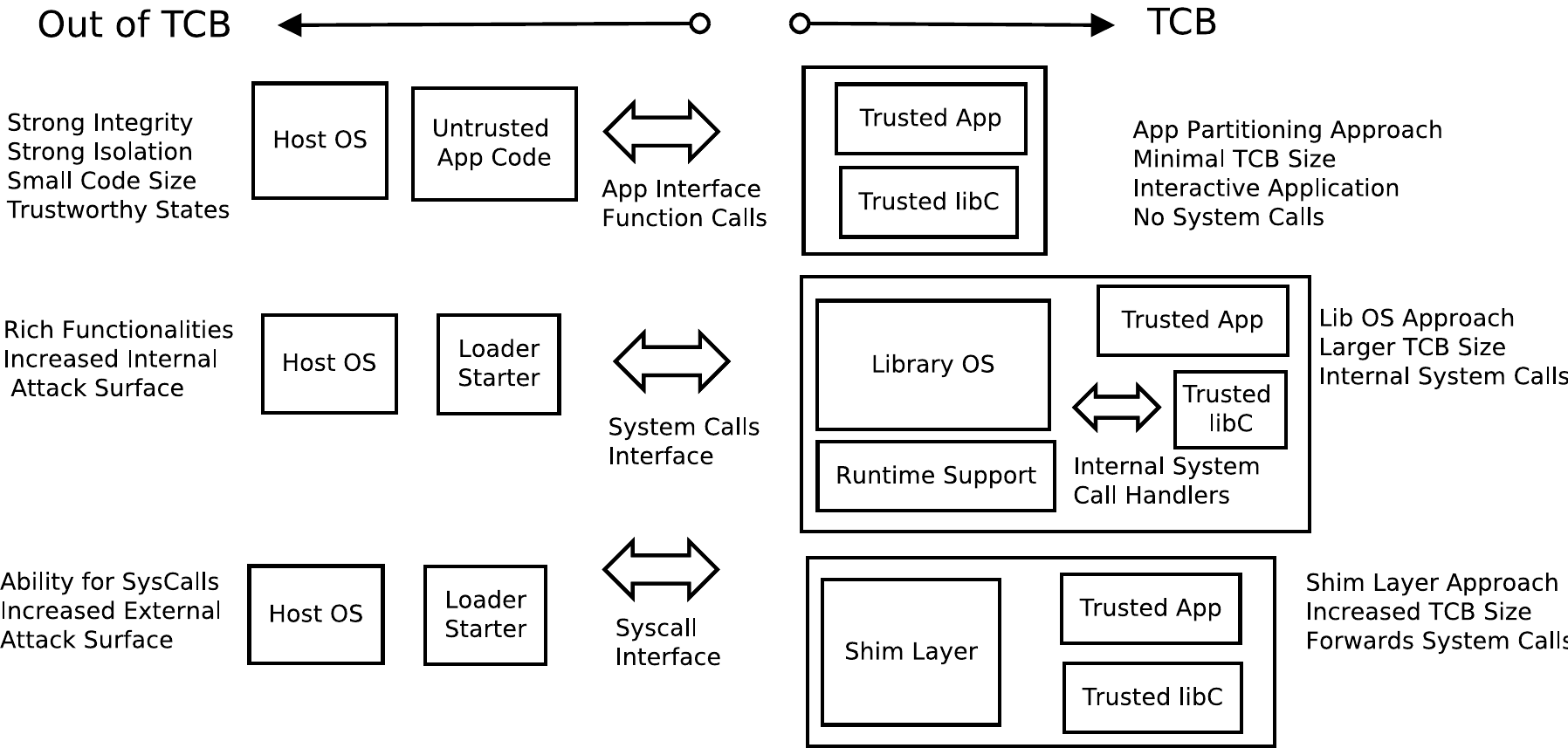}
  \caption{Overview of Development Models Security and Performance}
  \label{fig:development}
\end{figure*}

\textbf{Secure Loader:} A \emph{\acrlong{crc}} application may require a secure code loader for loading code into the trusted execution environment.
\acrshort{tpm} offers fixed functionalities for system software.
A separate software-based code loader in a system may be assisted by \acrshort{tpm}\@.
Trustzone developers must load the initial secure-world code themselves, and they can fetch other code later on.
They may place keys into the protected storage of Trustzone, which are required to load a secret code securely~\cite{ARM2009}.
However, this may not scale for multiple code providers to cooperate yet.
In the case of Intel's \acrshort{sgx}, untrusted components are free to load any software into the enclave, but attestation mechanisms allow users to verify the content of an enclave before proceeding with execution.
Attestation mechanisms utilise the binary hash of the enclave, compromising all secrecy of enclave code itself.
Initially, there was no secure loader for the enclaves.
Currently, Intel's \acrshort{sgx} SDK~\cite{SGXdevGuide21,SGXenclaveWritingGuide} offers integrated features as well as third-party solutions for secure code loaders. 
However, this raises other issues such as lack of property-based attestation and malicious secret behaviour in enclaves.

\textbf{Maturity:} A hardware candidate for \emph{\acrlong{crc}} should be mature enough to be a valid choice for commercial applications.
\acrshort{tpm} is the most accepted trusted hardware solution among current commodity systems, and it provides secure solutions for \acrlong{rts} and \acrlong{rtr}~\cite{tpmSpec}.
ARM Trustzone is another mature technology, widely deployed in most of the current mobile devices~\cite{Alves2004}.
At the time of writing, Intel's \acrshort{sgx} technology seems to require additional time to obtain the maturity necessary for enterprise-level solutions.

\textbf{Cost Efficiency:} The trusted hardware should be cheap and accessible for scalable deployment, considering the expected prevalence of IoT devices in the near future.
\acrshort{tpm} is one of the cheapest trusted hardware available in the market in terms of both hardware and development cost.
In contrast, ARM Trustzone-based hardware are more expensive on the market, and incur additional high licensing cost for production.
Intel's \acrshort{sgx} hardware is also relatively expensive, but its license allows for free and open-source development.

\begin{table} [b]
  \centering
  \resizebox{1\textwidth}{!}{%
  \begin{tabular}{c      |       c       |        c          |             c       |        c      |       c      }
    \toprule
                         \backslashbox{Type}{Feature}  &    TCB Size        &     Interface           &   OS Library      &   License   &  Loader      \\
    \midrule
    \shortstack{ Intel's SGX SDK }        &   Low    &     Interactive      &   Minimal        &    Open Source        & Plain  \\  \hline
    \shortstack{ GrapheneSGX SDK }        &  High    &     No              &    Full          &    Open Source         & Plain  \\   \hline
    \shortstack{ Shim Containers }        &  Medium  &     No              &    Shim Layer    &   Expensive           &  Plain  \\   \hline
    \shortstack{ Trustzone }              &  Low     &     Independent     &    No            &    Expensive          & Secure   \\   \hline
    \shortstack{ Ideal }                 &   Low     &     Independent      &   Minimal        &    Open Source      & Secure   \\

  \bottomrule
\end{tabular}
}
\caption{Comparing development model features of different types of SDKs/systems.}\label{compareDEV}
\end{table}

\subsection{Programming Models}
\label{sec:dev_domain}
In this section, we explain the programming models of secure applications.
We evaluate four types of development using different approaches.
Figure~\ref{fig:development} shows three approaches to enclave development with system support.
These approaches have different interfaces and they result in different \acrshort{tcb} sizes.
Interfaced applications can provide strong integrity and isolation, smaller code size and trustworthy states.
They follow the application partitioning approach: they can be interactive, and depending on the goals of the application the enclave may need to interact with the system in order to perform system calls or by design it may not require system calls.
Enclave applications with embedded LibOSes feature a larger \acrshort{tcb} size.
They can offer rich functionalities through the libOS support; however, the increased \acrshort{tcb} size increases the attack surface.
System calls can be kept inside the enclave.
The alternative approach to providing system support is the use of a shim layer, which also reduces the \acrshort{tcb} size.
This approach can provide a hybrid method for application development.
\acrshort{tcb} size can be kept relatively smaller than using a LibOS approach, but still provide a secure interface to the underlying system.
We show the development models in Table~\ref{compareDEV} and evaluate them by five metrics: (1) the \acrshort{tcb} size, (2) the communication interface of the enclave and the outside system, (3) the status of OS library support or use, (4) the status of development licenses (\eg~open development), (5) the code loader for initiating the execution.

The development models could be expanded on in a separate study with other SDKs such as Microsoft OpenEnclave~\cite{microsoft2020OpenEnclave} supporting multiple \acrshort{tee}s, SGX-LKL~\cite{priebe2019sgx} for \acrshort{sgx}, Rust-EDP~\cite{rust2020EDP} for \acrshort{sgx}, Google Asylo~\cite{google2020asylo} SDK, Keystone~\cite{lee2019keystone} SDK for RISC-V, Sancus~\cite{noorman2017sancus} SDK\@.
Bulck~\etal~explored the security vulnerabilities of these SDKs~\cite{van2019tale}.

\textbf{\acrshort{tcb} Size:} Intel \acrshort{sgx} SDK provides trusted libraries developed by Intel.
The \acrshort{tcb} size of the enclave is usually low. 
GrapheneSGX includes Library OS which adds substantial amount of code into the \acrshort{tcb}.
Shim Containers increase the \acrshort{tcb} size more than regular development with Intel \acrshort{sgx} SDK, but still less than a Library OS\@.

\textbf{Interface:} Intel \acrshort{sgx} SDK allows developers to define the interface of an enclave via \acrlong{edl}\@.
The enclaves can interact securely with untrusted components through the interface.
Enclave developers are responsible for designing and implementing a secure interface.
GrapheneSGX provides no interaction with untrusted components; its design principle requires strong isolation between unmodified applications and untrusted components of the system.
Applications can still make HTTP calls for communication with the outside world.

\textbf{OS Library:} Intel \acrshort{sgx} SDK contains only minimised trusted libraries developed by Intel.
GrapheneSGX SDK includes a full Library OS which helps any unmodified application execute inside their enclaves.
Shim Containers use Shim Layers to pass calls on from the trusted environment to the untrusted environment.

\textbf{License:} Both Intel \acrshort{sgx} SDK and GrapheneSGX SDK are free and open source.
Container solutions are usually not open-sourced and may incur high expenses on the side.

\textbf{Loader:} In the initial model, both Intel \acrshort{sgx} SDK and GrapheneSGX SDK load plain enclave code from the untrusted and inspectable environment. 
Thus, enclaves (\ie~the loaded code) are visible to the owner of the execution environment.
The loading mechanism in Intel's SDK has been updated to run with a \acrlong{pcl}, and recent research shows~\cite{kuccuk2019managing, silva2017dynsgx, bauman2018sgxelide} how the algorithms and enclave code can be protected.

\subsection{Attestation Mechanisms}\label{sec:att_domain}

An attestation mechanism provides evidence to be presented to verifying entities.
These pieces of evidence can consist of the hash of the binaries or any other means by which the summary of execution can be represented. 
The communication complexity of attestation schemes can cause an overhead between the participants.
We summarise different attestation mechanisms in Table~\ref{compareATT}, evaluated in six aspects: (1) what the evidence is based on, (2) how the evidence is affected by software updates, (3) the level of accuracy presented by the evidence, (4) the complexity of the communication flow, (5) whether code secrecy is maintained by the evidence, (6) whether the scheme can provide forward reliance with rules and policies.

\textbf{The Basis of Evidence:} In Diffie-Hellman and SIGMA Protocol-based remote attestation, the QUOTE signed by the \acrlong{qe} serves as evidence.
In TLS-based remote attestation, the communication is additionally simplified with certificates.
The property-based attestation mechanism uses features as a basis of the evidence.
Final State Attestation relies on the hash that was previously trusted by verifiers.
The local attestation relies on the REPORT generated by CPU firmware in an enclave itself.

\textbf{The Impact on Software Change:} If the software of the secure application changes in any way, presented evidence also changes in all attestation mechanisms except property-based attestation.
In property-based attestation, the evidence would only change if the feature it is based on undergoes a major change, otherwise it will remain under the trustworthiness threshold defined by a verifying party.

\begin{table}[!t]
  \centering
  \resizebox{1\textwidth}{!}{%
  \begin{tabular}{c       |    c       |        c          |             c       |        c      |       c     |    c  }
    \toprule
                     \backslashbox{Type}{Feature}                   &    Based On        &     Update           &   Accuracy     &   Flow   &  Secrecy  & Forward    \\
    \midrule
    \shortstack{ DH/SIGMA-based  \\ Remote Attestation }    &   Quote    &     Changes      &    High        &    Complex        & No   &  No \\  \hline
    \shortstack{ TLS-based Remote  \\ Attestation }         &   Cert.    &     Changes       &   High        &    Simplified    & No   &  No \\   \hline
    \shortstack{ Property-Based \\ Attestation }            &   Features &     Static       &    Custom        &    Simplified      & Yes  &  Yes  \\   \hline
    \shortstack{ Final State \\ Attestation  }              &   Hash      &    Changes      &    High        &    Simplified     & No   &  Yes  \\  \hline
    \shortstack{ Local \\ Attestation  }                       &   Report    &    Changes     &    High        &    Simplified   & Yes  &  No   \\

  \bottomrule
\end{tabular}
}
\caption{Comparing different aspects of Attestation Mechanisms.}
\label{compareATT}
\end{table}

\textbf{The Accuracy of Attestation:} Attestation evidence can tell little or more about attested software.
In property-based attestation, the accuracy depends on the algorithm that analyses the secret code.
High accuracy and secrecy are mutually exclusive notions, and a trade-off between them has to be made.
All other attestation mechanisms have high accuracy, as the evidence can give away the software version of the code used.

\textbf{The Communication Flow of Attestation:} The old type of remote attestation mechanism used by Intel \acrshort{sgx} SDK had a relatively complex communication flow with approximately 14 calls.
All other attestation mechanisms now offer simplified communication flow between the verifying and proving parties.

\textbf{The Secrecy for Attested Code:} Local attestation and property-based attestation can provide secrecy for the attested code.
Local attestation can prove that two parties are located on a local machine.
Based on a generated report, a verifying party can be informed that a prover is executing on the same hardware, without the proving party having to disclose the content of their assets.
In property-based attestation, a verifier obtains as much of the feature list as the prover permits without learning the source code of prover.
In other schemes, the verifier or another external party must approve that a certain piece of evidence belongs to a certain prover by replicating the same evidence generation method, for which they require delivery of the source code of a prover.

\textbf{The Future Reusability of Evidence:} In Final State Attestation, a verifier can rely on previously generated a piece of evidence, assuming that the prover does not change its state after the attestation.
In property-based attestation, similarly, a verifier can trust a prover without requiring any further approval after each update of the prover's code.
Once a verifier approves a list of features that the prover's code can access and perform, the prover can update its code as long as it satisfies the conditions of the verifier's previous approval.

\subsection{Participant Roles in Confidential Remote Computing}
\label{sec:prt_domain}

Table~\ref{tab:comparePRT} summarises the various roles of participants in \emph{\acrlong{crc}}.
These roles can be fulfilled either by a single or any number of participant entities, and any given entity can act in more than one role.
For example, in a blockchain-based system the hardware owner may utilise millions of devices owned by different participants, and combine their computational power to offer a unified infrastructure.
On the other hand, a homomorphic encryption scheme can represent a single hardware owner, operating with a single algorithm owner, but receiving input from multiple data owners.
In an ideal configuration, high numbers of these entities would need to be able to join and leave the computation at any point in time.

\begin{table}[!b]
  \centering
  \resizebox{\columnwidth}{!}{%
  \begin{tabular}{c                        |           c       |        c          |             c       }
    \toprule   \backslashbox{Role}{Feature}     &    Demands For    &     Assets    &   Threat Against       \\
    \midrule
    \shortstack{ Hardware Owner }            &   Revenue              &     Computation Platform        &    Data, Algorithm      \\   \hline
    \shortstack{ Data Owner  }               &    Results, Revenue      &    Private Data      &    Algorithm        \\  \hline
    \shortstack{ Algorithm Owner }           &    Results, Revenue  &    Secret Algorithm     &    Data            \\
  \bottomrule
\end{tabular}
}
\caption{Comparing Participant Roles}
\label{tab:comparePRT}
\end{table}

\textbf{What do they demand?} \emph{\acrlong{ho}s} (or hardware resource providers) demand to be able to make more revenue off their existing infrastructure.
The \emph{\acrlong{ao}s} need large data sets to run their algorithms and perform analysis.
They either demand the result of the computation for themselves, or they are interested in making revenue with the algorithm as a provided service.
\emph{Data Owners} request new algorithms to process their data from algorithm owners.
Like algorithm owners, they too are interested in computational results, or in making revenue by allowing the use of their data.
\emph{Data Owners} are also interested in computational results of other \emph{Data Owners}.

\textbf{What are their assets?} 
\emph{Data Owners} store large confidential data sets, they also protect the privacy of individuals in the data set.
\emph{\acrlong{ao}s} maintain a private algorithm they developed.
Such an algorithm is commercially valuable and requires secrecy.
\emph{\acrlong{ho}s} have a level of trustworthiness and reputation that they must maintain in order to maximise their revenue.

\textbf{What do they threaten?} 
\emph{Data Owners} and \emph{\acrlong{ho}s} are curious about the secret algorithm.
They pose a potential threat against the algorithm; they may for example collaborate to leak the secret algorithm.
\emph{\acrlong{ho}s} and \emph{\acrlong{ao}s} are curious about data sets.
They take any opportunity to threaten private data sets, and they may collaborate to diminish the privacy of individuals in data sets.
As the algorithm processes the data, it is easier for a \emph{\acrlong{ho}} to collaborate with the \emph{\acrlong{ao}} against the \emph{Data Owner} to leak or signal the private data.
In addition, \emph{\acrlong{ho}s} are also curious about both secret algorithms and private data.

\section{Trade-offs in working towards Ideal Confidential Remote Computation}
\label{sec:tradeoffsCRC}
\begin{figure}[htp!]
  \centering
  \includegraphics[width = 1\textwidth]{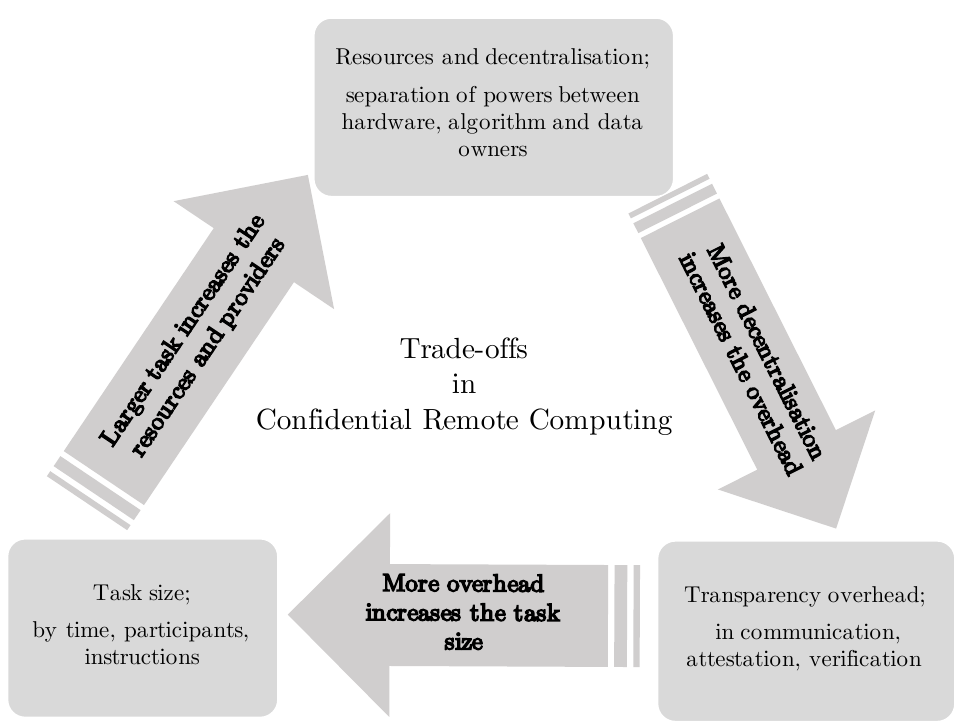}
  \caption{\emph{\acrlong{crc}} and its trade-offs.
 The initial requirements can begin from any of given node, are in turn complicated by the other parameters, increasing their cost. }
 \label{fig:crc-tradeoffs}
\end{figure}

We define three requirements for the ideal case.
Because these requirements are difficult to satisfy all at once, application developers usually have to make trade-offs according to their preferences.
A system might be defined by how much time its execution takes, the number of required participants to complete a task, and the rules or the instructions to operate it.
The task size then determines the amount of resources contributed by the resource providers. 
These resource providers are called hardware owners, data owners, and algorithm owners.
The providers can consist of multiple parties or a single entity.
If decentralisation is required in the system, this can increase the communication overhead between these entities.
Transparency can be provided through attestation, communication and verification processes. 
If more transparency is required to convince the participating entities, then this reduces the portion of the task that can be completed in a given time.
In Figure~\ref{fig:crc-tradeoffs}, we show that full decentralisation, complex task size, and full transparency are difficult to achieve at the same time. 
Each of these parameters complicates the next one and causes an additional overhead.
For example, a larger task size requires more resources.
The resources can be provided by a centralised authority, but a decentralised resource provider may not be able to fulfil the demand in restrictions.
A real-world example can be that visa cards can handle a high volume of transactions centrally, but block-chain technologies cannot reach the same level of throughput owing to their decentralised nature.
The second challenge is that high decentralisation makes transparency operations even more difficult.
However, it is required in order to safeguard decentralised activities with transparency, preventing creeping centralisation.
Transparency then increases the overheads in communication, attestation or verification processes.
The accumulation of these overheads render the completion of a task effectively impossible.

\subsection{Control Decentralisation of Computations}

Today's cloud services retain and accumulate their control of computations and become centralised powers. 
Data and  algorithms are not hidden from most cloud providers.
The first step to break the authoritarian execution models of the cloud is to keep data confidential in-use to answer increasing privacy concerns.
For a better decentralisation, the computations must be independent of the underlying hardware pool.
While manufacturers do not control what is computed using their devices, most of the cloud hardware owners are selective on what can be computed.
Hardware owners must remain neutral, lest they collude with either data or algorithm owners in alternative channel attacks on each other.

\subsection{Scalability under Limitations of Time, Participants and Instructions}

Ceteris paribus, a cloud system offering no security may scale higher at a lesser cost.
One way to quantify scalability is the number of participants being served in an hour.
Cryptographic protocols can offer formal security properties, however, its performance is limited by the number of participants and operations.
In a fully scalable setting, systems must be able to support multiple hardware, data and algorithm owners having a joint computation.
The number of distinct participants may go up to millions or billions with general-purpose computation abilities.
Participants must be able to enter and exit the computations dynamically.

\subsection{Communication Overhead in Attestation Methods}
Automating the attestation mechanisms can help to reduce communication overheads. 
For example, a frequently updated system with binary attestation can slow down the entire process.
Attestation reports can be verified asynchronously, or attestation quotes can be reused when possible, the number of properties of the report can have variables to tolerate certain changes.

\subsection{Challenges on Keeping TCB Minimal}

While decentralising control, allowing for a high number of participants and automating the attestation process, the whole system should have the minimal possible trusted computing base.
An often repeated mistake is that \acrshort{tcb} size goes dramatically high, up to hundreds of thousands (or even millions) of lines of code.
This causes difficulties in formalisation and often results in an obscure failure of security analyses.





\section{Conclusions \& Insights Gained}

In this section, we provide a list of the lessons learned through \emph{\acrlong{crc}} model.
We also list some of the outstanding questions and aspects of system design requiring more attention.

\begin{figure}[htp]
  \centering
  \includegraphics[width = 1\textwidth]{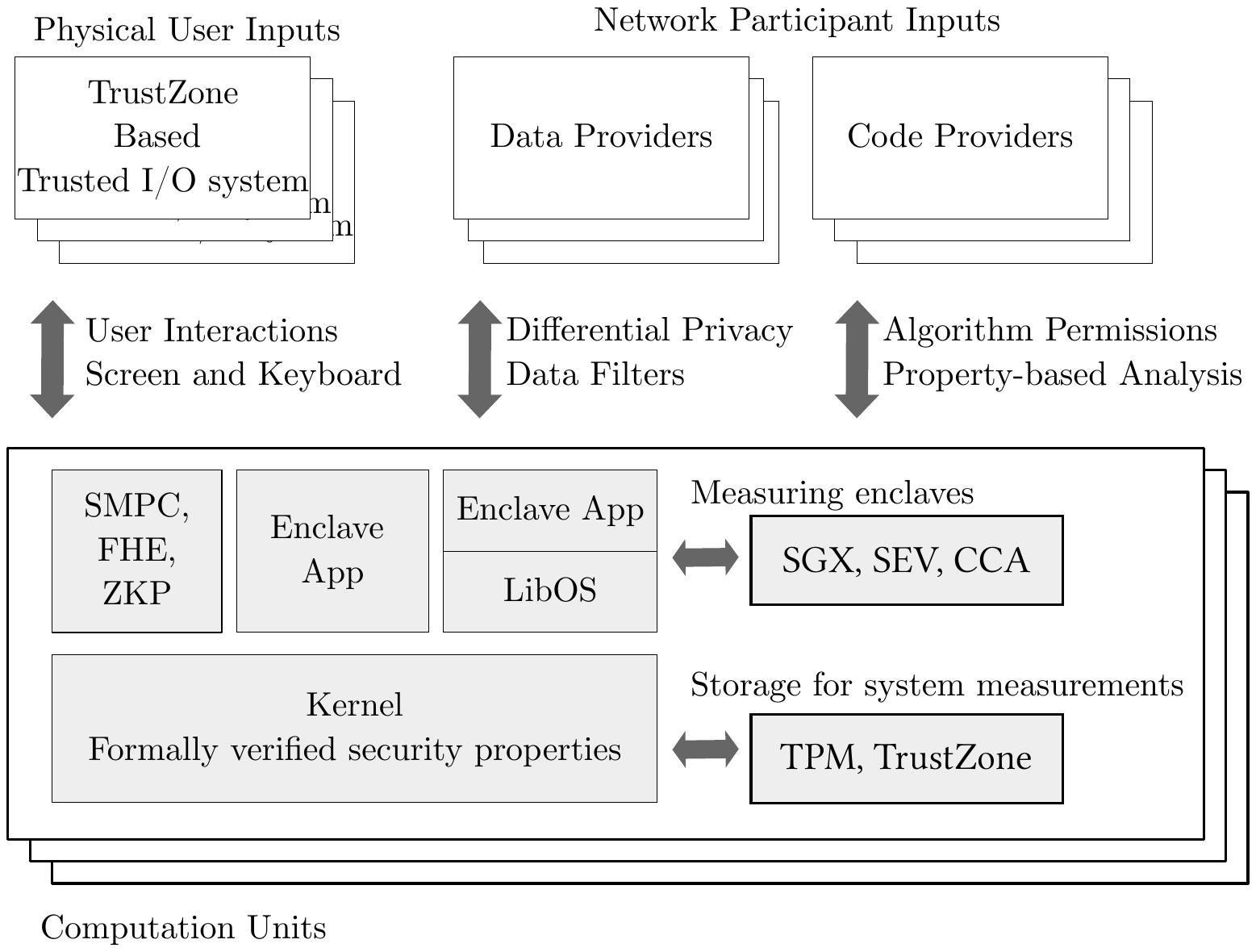}
  \caption{A unified architecture for \emph{Confidential Remote Computing} systems. 
  }
  \label{fig:unified-architecture}
\end{figure}

\subsection{The Unified Computing Model}
\label{sec:unifiedComputingModel}
We use the lessons learned to build a unified architecture for \emph{\acrlong{crc}} units, shown in Figure~\ref{fig:unified-architecture}.
Multiple hardware technologies and techniques are integrated into the ideal setting.
A micro-kernel may not necessarily improve the security of the system by itself. 
Still, a formally verified micro-kernel for this architecture can improve the security of the overall system (one candidate for such a kernel is seL4~\cite{klein2009sel4}).
A subset of the model can be used in smart grid systems, in e-voting systems or in hub/gateway systems of the Internet of Things (IoT).
End-to-end cryptography can be used in isolated enclave applications. 
Enclave code can also contain applications utilising zero-knowledge proofs (ZKP), secure multi-party computation (SMPC) and fully homomorphic encryption (FHE) schemes.
Enclave measurement might be done by \acrshort{sgx} instructions, SEV instructions, ARM's CCA or future instructions.

\subsection{Lesson I: Underlying insecure system components enabling side-channel attacks}
In case of \acrshort{tee}s, many of the software-based side channel attacks become possible due to architectural choices. 
The key lesson here is that when designing enclaves and \acrshort{tee} applications, manufacturers initially claimed that OS can be fully insecure and enclaves will be still isolated and fully protected from the malicious OS, we learned that this is not possible in practice, and the importance of the underlying insecure system components should not be ignored.
Fully adversarial systems can deploy side-channel attacks against the applications running on top of them.
Hardening techniques can mitigate side-channel attacks, but in order to avert the risk completely it would be necessary to remove the fundamental conditions inviting them.
A system supporting enclave execution can be built with formally verified kernels (\eg~seL4) and utilise measured and attested boot features with \acrshort{tpm}.
This can ensure that the enclaves operate in benign and trusted systems.
Electronic voting systems and IoT gateway systems managing a high number of entities and deploying mission-critical operations can benefit from such secure structures.

\subsection{Lesson II: Algorithm status and physical location for security}

One of the requirements of software-based side-channel attacks targeting the enclave software from an insecure operating system is having the enclave code (algorithm) publicly available for the attacker.
Hiding the algorithms through dynamic code loaders at run-time can help reduce leakages from attacks.
Nevertheless, keeping algorithms hidden is not sufficient by itself to entirely avert side-channel attacks.
The key point to somewhat mitigate attacks is that the algorithms must be hidden from the hardware owner. 
If a malicious admin or hardware owner has access to the enclave code, this can help them deploy sophisticated attacks more easily.
Furthermore, if potentially malicious hardware and algorithm owners collude, they could develop covert channels to signal and leak the data.
This collusion could then invalidate the security guarantees offered by the underlying secure hardware.
Therefore, enclave developers must be independent of the infrastructure providers, why we say \emph{must be} independent is because the collusion would ruin the security as we discussed in Section~\ref{sec:tradeoffsCRC} and~\cite{kuccuk2019managing}.
Suppose a company (or an entity) claims to provide both enclave software and secure hardware resources. 
In that case, this may lead to a high risk of data signalling due to explicit collusion, even if the enclave code is presented to or attested by the data owners.
As a countermeasure, the whole system must be attested as designed with secure kernels, rather than attesting only the enclave code.
Moving the physical trusted execution environment to the data provider's environment and deploying physical network restrictions can help in terms of confidentiality; however, such a system setting would face the scalability challenges of multi-party computation solutions.
This has been a major insight because when the \acrshort{tee}s were initially offered in commodity hardware, the researchers and manufacturers simply did not know enough to define the right threat model.

\subsection{Lesson III: Role of enclaves and verification in secure computing}

Enclave technology can help verify remotely that a given piece of code purported to have run indeed did run.
In this manner, trusted hardware generates evidence for an attestation process.
However, the actual behaviour of the code, whether it matches its expected security properties, is a question for verification technologies.
In other words, formal verification technology helps prove that a piece of code does what it is supposed to do.
These two steps are ideally both necessary and not here to replace one another.
While designing the enclave applications, partitioning can give enclave developers an opportunity to enable verification of the code base.
For example, migrating the security-critical parts into the isolated memory can allow this.
Design choices, unnecessarily large code base, and poor partitioning would ruin this chance.
Interested readers are referred to the earlier publications~\cite{Kucuk2016},~\cite{kuccuk2019managing}, and Section~\ref{sec:xchart}.

\subsection{Lesson IV: Potential issues with a model suggesting to initialise enclaves first in a system} 
What should we expect to happen if enclaves were initialised independently of the host computer, before the system starts?
Parallel to the model of SMM, which inadvertently yielded highest system privileges to malicious actors, if enclaves were to be initialised first, they could be expected to draw similar attacks~\cite{wojtczuk2009attacking2}.
Nevertheless, \acrshort{sgx} memory for enclaves is in fact allocated by default at the earliest stage of the system start.
The firmware knows in advance how much memory will be given for \acrshort{sgx} enclaves, but the content of enclaves is not persistent and binaries are loaded later.
That is why early initialisation of \acrshort{sgx} memory does not cause similar problems as it used to with SMM.

In contrast to its characteristics in \acrshort{sgx} enclaves, content in a Trustzone secure world remains persistent.
Trustzone applications are initialised first in the system, but they can trigger the system to take control on demand via interrupts, for example to protect their integrity.
Trustzone promises to mitigate potential exploits compromising its persistent content via secure \acrshort{tcb} updates.

\subsection{Lesson V: There are now Multiple Root(s) of Trust in a system}
A novel use of the \acrshort{txt}/\acrshort{drtm}, as introduced in Flicker~\cite{McCune2008}, evolved into the use case of hardware enclaves with \acrshort{sgx}.
Multiple measurements in a system are stored in different \acrlong{pcr}. 
Incorrect use of \acrshort{pcr} already has difficulties representing the system state correctly.
With \acrshort{sgx} hardware implementation and a \acrshort{tpm} in a system, an additional problem arises alongside that of wrong use of \acrshort{pcr} values.
This development places the important new responsibility to securely combine \acrlong{rtm} on system designers and security architects.
Use of \acrshort{sgx} in production without merging its \acrlong{rtm} with that of \acrshort{tpm}-enabled systems threatens to fork the \acrlong{rtm} concept into two independent measurement mechanisms.
This can be avoided via using the same trusted firmware on the bottom in future hardware designs.
For example, when ARM or RISC-V introduces enclaves similar to \acrshort{sgx} for user level applications, they may use the same \emph{trusted firmware} to manage the enclaves and the Trust-Zone.

\subsection{Lesson VI: TCB minimisation should not be neglected}
Researchers put library operating systems with hundreds of thousands of lines of code inside enclaves.
This creates a large and insecure software stack.
To avoid such an outcome, enclaves must be kept as minimal as possible.
LibOSes can boost functionalities of enclaves, but sacrifice the security guarantees derived from trusted hardware.
In-enclave OS support introduces the same security risks of an OS in the untrusted world.
The \acrshort{tcb} will be potentially vulnerable to attacks unless the new LibOS implementations give formal security guarantees.
Otherwise, enclaves with LibOSes may not be able to patch their systems, which are generally insecure due to being built upon a weak kernel ring structure (Ring 0 and 3).
It would be preferable to keep enclaves clean from the clutter of unnecessary, untrusted code.

\section{Future Work}
In the upcoming years, we anticipate that consumer devices (ARM/RISC-V) will be equipped with new user-level realms (similar to enclaves).
These devices' wide availability, mobility and decentralised setting can enable a new era of inter-connected, remote computations.
Thanks to the rich commodity software stack and libraries available for ARM architecture today, system designers and developers may offer \acrlong{crc} applications for consumer devices.
We encourage developers to work on solutions from e-voting systems to decentralised financial applications and data or algorithm rental services independent of centralised third-party servers. 
Towards a more decentralised setting for computations, we may see legitimate \textbf{digital embassies} of individuals running on devices of other consumers.

\subsection{Configuration Security, Modularity and Composition}
In complex systems, the configuration of systems with modular structures remains an open research challenge.
Security properties of sub-modules alone cannot serve as a representation of the properties of the entire composed system.
Formally verified kernels may be difficult to keep up to date in the frequently-changing (or update-receiving) world of general-purpose systems.
On the other hand, relatively smaller industrial-embedded systems and IoT devices can benefit from formal guarantees.

\subsection{Using Multiple Trusted-hardware Chips on a Single Consumer Device}
Taking measurements (\eg~checksum of code and data) in order to understand the trust status of a system is generally accepted as a useful strategy.
Today, a single commodity computer can have both \acrshort{tpm} and \acrshort{sgx} enabled.
This may potentially mean having two different \acrlong{rtm} (platform firmware and CPU xucode) by which to describe the system and represent its trustworthiness or lack thereof.
This poses the question which one or both of the two are to be actively used in representing a whole system, and in the latter case how their differential information output should be combined.
In order to evaluate the trustworthiness of a system, reports, quotes and evidences must be securely composed and presented in a unified form.
Using the right mechanism, for example the \emph{trusted firmware} for the right purpose in secure composition remains the crucial challenge.

\section*{Acknowledgements}
Some of this work is supported by the InnovateUK ManySecured project.
We thank Sean Smith, O Yaman, A Acar, M Geden, I Heinemann, for their helpful discussions and reviews.




\bibliographystyle{unsrt}
\bibliography{refs}

\end{document}